# INTERNAL ELECTRIC FIELD ORIGINATED FROM MISMATCH EFFECT AND ITS INFLUENCE ON FERROELECTRIC THIN FILM PROPERTIES

M.D.Glinchuk, A.N.Morozovska

Institute for Problems of Materials Science, National Academy
of Science of Ukraine, Krjijanovskogo 3, 03142 Kiev, Ukraine

**Abstract**

The ferroelectric thin film properties were calculated in phenomenological theory framework. Surface energy that defined boundary conditions for Euler-Lagrange differential equation was written as surface tension energy. The latter was expressed via surface polarization and tension tensor related to mismatch of a substrate and a film lattice constants and thermal expansion coefficients.

The calculations of the film polarization distribution, temperature, thickness and external electric field dependence and hysteresis loops as well as average dielectric susceptibility dependence on temperature and film thickness have been performed allowing for mismatch-induced polarization $P_m$, leading to appearance of internal thickness dependent field. It has been shown that this field influences drastically all the properties behaviour. In particular the polarization profile becomes asymmetrical, average polarization temperature dependence resembles the one in the external electric field, and there is possibility of external field screening by the internal one. The obtained asymmetry of hysteresis loop makes it possible to suppose that the self-polarization phenomenon recently observed in some films is related to mismatch effect. The thickness induced ferroelectric-paraelectric phase transition has been shown to exist when the $P_m$ value is smaller than the polarization $P_S$ in the bulk, i.e. at $|P_m|/P_S < 1$. The large enough mismatch effect ($|P_m|/P_S > 1$) could be the physical reason of ferroelectric phase conservation in ultrathin film. The possibility to observe the peculiarities of the films properties temperature and thickness dependencies related to mismatch effect is discussed.

Keywords: surface tension, thin film, mismatch effect, phase transition.

PACS: 77.80.-e, 77.55.+f, 68.03.Cd, 68.35.Gy,

## 1. INTRODUCTION

The influence of substrate on ferroelectric thin film properties is known to be related with the appearance of mechanical tension due to mismatch between lattice constants and thermal expansion coefficients of the film and its substrate as well as growth imperfections. This phenomenon was taken into account by consideration of the uniform mechanical tension tensor $u_{ii}$ ($i = x$, $y$, when $z$ is the normal direction to the surface). Inclusion of this tensor into a free energy of bulk ferroelectrics, allowing for its coupling to electric polarization via electrostriction in cubic lattices and subsequent minimization of the free energy has shown [1], [2] the lowering of the cubic to tetragonal symmetry as one can expect for any film and renormalization of free energy coefficients and so ferroelectric phase transition temperature. The obtained results are characteristic for the bulk ferroelectrics under uniform mechanical tension of the considered symmetry. The obtained in [1], [2] effects were completely independent on the film thickness and so they did not contribute to the size effects of the film properties. The latter are known to be related to the difference between surface and bulk properties, so that gradient of polarization appears in free energy functional, which after variation gives Euler-Lagrange differential equation with boundary conditions related to surface energy parameters (see e.g. [3] and ref. therein). It is obvious that mechanical tension related to the mismatch effects can contribute to surface energy and so to the boundary conditions and finally to all the properties of the film. On the other hand the appearance of misfit dislocations at some critical distance from the film surface, when the appearance of the dislocations become energetically preferable can decrease essentially the mechanical tension inside the thick enough film allowing for the critical thickness for dislocation appearance is several tens or hundred nm [4], [5]. Therefore for the films thinner than the critical thickness of misfit dislocations appearance the mechanical tension can be considered as uniform while for the thicker films it transforms into smaller non-uniform tension because of misfit dislocations influence. The authors of [4], [5] propose to consider the latter effect approximately by substitution of some smaller effective tension for the $u_{ii}$ value allowing for that for thick enough films the mechanical tension becomes negligible small.

It is obvious that mechanical tension effect is the most important for thin enough ferroelectric film and it has to be included both into the surface energy and into the "bulk" part of free energy functional. In the present work we performed for the first time the calculations of dielectric permittivity, characteristics of thickness induced ferroelectric-paraelectric transition, the conditions of its absence, self-polarization and hysteresis loops by solution of Euler-Lagrange equation for inhomogeneous polarization with boundary conditions which include mechanical deformation contribution caused by mismatch effect.

## 2. MODEL AND GENERAL FORMALISM

### 2.1. Free energy functional

Let us consider ferroelectric thin film with the thickness $l$ $\left(-l/2 \leq z \leq l/2\right)$ and polarization in the direction z normal to the surface ($P_z \equiv P$). In phenomenological theory approach free energy functional can be written in the form [3]:

$$\Delta G = G - G_0 = \int g_v dv + \int g_s ds . \qquad (1)$$

Here the first and the second integral reflect polarization dependent contribution of the bulk and the surface of the film, $G_0$ is polarization independent part of the free energy.

The bulk free energy density $g_v$ can be represented as:

$$g_v = \frac{\alpha P^2}{2} + \frac{\beta P^4}{4} + \frac{\delta}{2}\left(\frac{dP}{dz}\right)^2 - P\left(E_0 + \frac{E_d}{2}\right) \qquad (2)$$

Here the coefficients $\alpha$, $\beta$ are supposed to be renormalized by mechanical tension on the base of procedure developed in [1]; $E_0$ is external electric field; $E_d$ is depolarization field, its value for the case of single-domain insulator film with super-conducting electrodes can be written in the form [6]:

$$E_d = 4\pi\left(\overline{P} - P\right). \qquad (3)$$

Hereafter the bar over a letter denoting a physical quantity represents the spatial averaging over the film thickness. It is seen that in the bulk samples with homogeneous polarization $\overline{P} = P$, depolarisation field is absent ($E_d = 0$), while in the thin films with inhomogeneous polarization $E_d \neq 0$. It should be noted, that the Eq. (2) is suitable for the description of the second order phase transitions with $\beta > 0$.

Allowing for surface energy is related to surface tension [7], it is possible to represent the second term in Eq. (1) similarly to [8]:

$$G_s = \sum_{i=1}^{2} \int \mu_i u_{xx}^{(i)} u_{yy}^{(i)} dx dy . \qquad (4)$$

Here $\mu$ and $u_{jj}$ ($j = x$, $y$) are respectively the surface tension coefficient and strain tensor components, $i = 1$, 2 reflects the contribution of the film two surfaces.

In what follows we will consider two main contributions to the deformation tensor, namely the first considered in [8] related to surface polarization $P_s$ via piezoelectric effect that exists even in a cubic symmetry lattice near the film surface, and the second is related to mismatch effects discussed in the introduction. Therefore

$$u_{xx}^{(i)} = u_{xxm}^{(i)} + d_{xxz}^{(i)} P_{zi}^{(i)}, \quad u_{yy}^{(i)} = u_{yym}^{(i)} + d_{yyz}^{(i)} P_{zi}^{(i)}, \qquad (5)$$

where $z_1 = l/2$, $z_2 = -l/2$, $d_{jjz}$ is the coefficient of piezoelectric effect, $u_{jjm}$ is the tensor of mechanical strain that is proportional to the difference in the lattice constants and thermal expansion coefficients between a substrate and a film as well as to the growth imperfections. In what follows we will consider an epitaxial film with bulk cubic symmetry, e.g. BaTiO$_3$, PbTiO$_3$ etc. In such case $u_{xx}^{(i)} = u_{yy}^{(i)}$ because $d_{xxz}^{(i)} = d_{yyz}^{(i)} \equiv d^{(i)}$, $u_{xxm}^{(i)} = u_{yym}^{(i)} \equiv u_m^{(i)}$ and so the product $u_{xx}^{(i)} u_{yy}^{(i)}$ in Eq. (4) can be rewritten as $u_{xx}^{(i)2} = u_m^{(i)2} + 2d^{(i)} u_m^{(i)} P_{zi} + d^{(i)2} P_{zi}^2$, where the first term is independent on polarization, while the last two terms are defined the surface free energy. Allowing for Eqs. (2), (3), (4), (5) the free energy (1) acquires the form:

$$\Delta G(P) = \frac{1}{l} \int_{-l/2}^{l/2} dz \left[ \frac{\alpha}{2} P^2(z) + \frac{\beta}{4} P^4(z) + \frac{\delta}{2} \left( \frac{dP(z)}{dz} \right)^2 - P(z) E_0 \right] +$$

$$+ \frac{2\pi}{l} \int_{-l/2}^{l/2} dz \left( P(z) - \overline{P} \right)^2 + \frac{\delta}{2l} \left[ \frac{\left( P(l/2) + P_m^{(1)} \right)^2}{\lambda_1} + \frac{\left( P(-l/2) + P_m^{(2)} \right)^2}{\lambda_2} \right] \tag{6}$$

Hereinafter

$$\lambda_{1,2} = \frac{\delta}{2\mu_{1,2} d^2}, \quad P_m^{(1,2)} \equiv \frac{u_m^{(1,2)}}{d}. \tag{7}$$

Since the signs of $u_m$ and $d$ can be positive or negative both $P_m > 0$ and $P_m < 0$ are expected. Because of this we will choose $P_m$ sign by comparison with experiment.

The renormalized coefficient $\alpha$ in (6) has the form [1]:

$$\alpha(T) = \alpha_T \left( T - T_c^* \right), \quad T_c^* = T_c + \frac{2 Q_{12} u_m}{\alpha_T (S_{11} + S_{12})}. \tag{8}$$

Here $T_c$, $\alpha_T$, $Q_{12}$ and $S_{11}$, $S_{12}$ are respectively ferroelectric transition temperature, inverse Curie constant, electrostriction coefficient and elastic modulus of the bulk material. It follows from expression for $T_c^*$ in Eq. (8) that for the given pair substrate-film it can be $T_c^* < T_c$ for $u_m > 0$ or $T_c^* > T_c$ for $u_m < 0$, because $Q_{12} < 0$ for the materials with perovskite structure [9]. It should be noted, that $T_c^*$ corresponds to ferroelectric-paraelectric transition temperature of bulk material under compressive or tensile external pressure of the considered symmetry.

## 2.2. Euler-Lagrange equation and boundary conditions

The equation for calculation of the polarization can be obtained by variation over polarization of free energy functional (6). This yields the following Euler-Lagrange equations and the boundary conditions:

$$\alpha P + \beta P^3 - \delta \frac{d^2 P}{dz^2} = E_0 + 4\pi (\overline{P} - P), \tag{9}$$

$$\left( P + \lambda_1 \frac{dP}{dz} \right) \Big|_{z = l/2} = -P_m^{(1)}, \quad \left( P - \lambda_2 \frac{dP}{dz} \right) \Big|_{z = -l/2} = -P_m^{(2)}. \tag{10}$$

In what follows we will consider the realistic situation of the film on the substrate with free-standing upper surface, where $u_m^{(1)} = 0$ and so $P_m^{(1)} = 0$. To find out the transition temperature in the ferroelectric film one has to solve the Eq. (9) with boundary conditions (10) at $P_m^{(1)} = 0$ and $P_m^{(2)} = P_m \neq 0$ (see Fig. 1).

Looking for the possibility to obtain the most clear analytical results, let us solve the linearized Eq.(9) with equal extrapolation lengths $\lambda_1 = \lambda_2 = \lambda$, namely

$$\alpha(T) P - \delta \frac{d^2 P}{dz^2} = E_0 + 4\pi (\overline{P} - P), \tag{11a}$$

$$\left(P + \lambda \frac{dP}{dz}\right)\bigg|_{z=l/2} = 0, \quad \left(P - \lambda \frac{dP}{dz}\right)\bigg|_{z=-l/2} = -P_m. \tag{11b}$$

The solution of (11) has the following form:

$$P_{Lin}(z) = \frac{E_0 - P_m \Psi(l)/2}{\alpha(T) + \Psi(l)}[1 - \varphi(z)] - \frac{P_m}{2}[\varphi(z) - \xi(z)]. \tag{12}$$

Hereinafter we used the following designations

$$\Psi(l) = 4\pi\left(\frac{2l_d}{l}\right)\frac{th(l/2l_d)}{1 + (\lambda/l_d)th(l/2l_d)}, \tag{13a}$$

$$\varphi(z) = \frac{ch(z/l_d)}{ch(l/2l_d) + (\lambda/l_d)sh(l/2l_d)}, \tag{13b}$$

$$\xi(z) = \frac{sh(z/l_d)}{sh(l/2l_d) + (\lambda/l_d)ch(l/2l_d)}. \tag{13c}$$

The characteristic length is $l_d = \sqrt{\delta/(4\pi + \alpha)}$. The expression for $\bar{P}$ could be calculated after taking into account that $\bar{\varphi} = \Psi(l)/4\pi$, $\quad \bar{\xi} = 0$, which leads to

$$\bar{P}_{Lin}(T,l) = \frac{E_0}{\alpha(T) + \Psi(l)}\left[1 - \frac{\Psi(l)}{4\pi}\right] - \frac{P_m\Psi(l)}{2(\alpha(T) + \Psi(l))}\left[1 + \frac{\alpha}{4\pi}\right]. \tag{14}$$

Notice, that $\bar{P}_{Lin}(T,l)$ diverges at the critical point $\alpha(T) + \Psi(l) = 0$ which is independent on $P_m$. It was shown earlier in [10], [11], that the origin of this divergence for the films with $P_m = 0$ is the thickness induced ferroelectric-paraelectric phase transition, $\bar{P}_{Lin}(T,l)$ being paraelectric phase polarization induced by external field $E_0$. The latter is not true in the considered case $P_m \neq 0$ as it follows from Eq.(14), so that the divergence of $\bar{P}_{Lin}(T,l)$ cannot indicate the phase transition point. As a matter of fact the second term in Eq.(14) could be considered as a film self-polarization originated from mismatch effect. In order to show this let us make some simplifications in Eq.(14), taking into account that for the most of ferroelectrics $l_d \approx \sqrt{\delta/4\pi} \sim 1 \div 10 \overset{\circ}{A}$ [9] and

$$l \gg l_d, \ \lambda \gg l_d, \ \alpha/4\pi \ll 1, \tag{15}$$

so the expression (14) could rewritten as

$$\bar{P}_{Lin}(T,l,E_0) \approx \frac{E_0 - P_m\Psi(l)/2}{\alpha(T) + \Psi(l)}. \tag{16}$$

It is clear from (16) that the combination $(E_0 - P_m\Psi(l)/2)$ plays the role of the effective field that determines the polarization amplitude. Therefore one have to apply the non-zero external field $E_0 \approx P_m\Psi(l)/2$ in order to compensate the internal self-polarization induced by mismatch effect. Possible experimental manifestation of this effect we shall discuss later.

Note, that the linearization of (9) is valid only for the small polarization amplitudes $|\bar{P}_{Lin}/P_S| \ll 1$ (hereinafter $P_S = \sqrt{\alpha_T T_C^*/\beta}$ is the spontaneous polarization of bulk material at $T = 0$ K), so that the obtained results (12), (14) could not be used in the vicinity of the point $T = T_C^*$ where the nonlinear term in Eq.(9) must be taken into account.

### 2.3. Free energy with renormalized coefficients

Let us find the approximate solution of the nonlinear Eq.(9) by the direct variational method [12]. We will choose the one-parametric trial function in the form of linearized solution (12) that satisfies the boundary conditions (10), namely:

$$P(z) = P_V[1 - \varphi(z)] - \frac{P_m}{2}[\varphi(z) - \xi(z)] \tag{17}$$

The amplitude $P_V$ must be determined by the minimisation of the free energy (6). Integration in the expression (6) with the trial function (17) leads to the following form of the free energy:

$$\Delta G(P_V) = \frac{A_m}{2} P_V^2 + \frac{D_m}{3} P_V^3 + \frac{B_m}{4} P_V^4 - P_V \left[ E_0 - E_m \right]. \qquad (18)$$

Under the validity of the inequalities (15) the renormalized coefficients in (18) have the following form

$$A_m(T,l) = 4\pi\theta \left( \frac{T}{T_C^*} - 1 + \frac{1}{\theta \, h(1+\Lambda)} + \frac{P_m^2}{P_S^2} \frac{(1+3\Lambda)}{4h(1+\Lambda)^3} \right), \qquad (19)$$

$$B_m(l) = \beta \left( 1 - \frac{11 + 27\Lambda + 18\Lambda^2}{6h(1+\Lambda)^3} \right), \qquad (20)$$

$$D_m(l) = -\frac{\beta P_m}{2h} \left( 1 - \frac{\Lambda^3}{(1+\Lambda)^3} \right), \qquad (21)$$

$$E_m(l) = E_S \frac{P_m/P_S}{2h(1+\Lambda)} \left( \frac{1}{\theta} - \frac{P_m^2/P_S^2}{3(1+\Lambda)^2} \right). \qquad (22)$$

Here the following designation are introduced:

$$\theta = \frac{\alpha_T T_C^*}{4\pi}, \qquad h = \frac{l}{2 \, l_d}, \quad \Lambda = \frac{\lambda}{l_d}, \qquad E_S = \alpha_T T_C^* P_S, \qquad P_S = \sqrt{\alpha_T T_C^* / \beta} \ . \qquad (23)$$

Note, that odd power $P_V^3$ in Eq.(18) is unusual for cubic symmetry perovskite structure ferroelectrics, this term as well as $E_m(l)$ are absent at $P_m = 0$ (see Eqs.(21), (22)), i.e. it is related to mismatch effect.

## 3. POLARIZATION AND HYSTERESIS LOOPS

The main advantage of free energy (18) is the fact that it is algebraic function of $P_V$, and thus the dependence of $P_V(E_0,T,l)$, can be derived directly from the minimisation of free energy (18) over $P_V$, namely

$$A_m(T, \, l)P_V + B_m(l)P_V^3 + D_m(l)P_V^2 = E_0 - E_m(l) \ . \qquad (24)$$

As an example the zero-field temperature dependence of polarization $P_V(T, E_0 = 0)$ is depicted in Fig.2, where the curves 1 correspond to the condition $E_m = 0$. Temperature dependence represented by curves 1 corresponds to those obtained in [10], [11] and $P_V = 0$ at the temperature of thickness induced ferroelectric-paraelectric phase transition $T = T_{cl}(l)$. The value of $T_{cl}(l)$ decreases with the film thickness $h$ decreasing and $T_{cl}(l) = 0$ at $h \le 8$, that corresponds to the critical thickness. At $P_m \ne 0$, $E_0 = 0$ the behaviour of $P_V(T, E_0 = 0)$ looks like that in the external field and for the thickest film ($h = 40$) it resembles the behaviour in the bulk samples (see e.g. [9]). To find out physical meaning of polarization $P_V$ let us calculate average polarization. In accordance with (13b), (17), (19) and (23) the average polarization acquires the form

$$\overline{P} = P_V \left[ 1 - \frac{1}{h(1+\Lambda)} \right] - \frac{P_m}{2} \frac{1}{h(1+\Lambda)} \qquad . \qquad (25)$$

Keeping in mind, that for the most of thin ferroelectric films $h \ge 10$, $\Lambda \ge 10$ one can see that $\overline{P} \approx P_V$.

The distribution of the ratio $|P(z)|/P_S$ over coordinate $z$ is represented in Fig.3 for several values of $P_m/P_S$ and $\lambda/l_d = 10$, allowing for the expressions (13b) and (13c) for $\varphi(z)$ and $\xi(z)$ respectively. One can see that at $|P_m/P_S| < 1$ the profiles $P(z)$ looks like those obtained earlier for $P_m = 0$ [10], [11], and at $|P_m/P_S| > 1$ the strong asymmetry arises. It originates from the asymmetry of boundary conditions (11b) caused by the presence of a substrate.

The typical forms of hysteresis loops $\bar{P}(E_0)$ are represented in Fig.4a. As it follows from the expression (24) and Fig.4a the quasi-equilibrium hysteresis loop shifts as a whole along $E_0/E_c$ axis under $P_m/P_S$ increasing. Right hand side shift of the hysteresis loop corresponds the experimentally observed one [13] in the self-polarized film (compare the form of the curve 2 for $h = 40$ in Fig.4a with the loop in Fig.4b). Indeed the authors of [13] came to the conclusion, that the shift of the loop is attributed to the natural polarization more than 50 $\mu C/cm^2$ without poling treatment. Because the direction of the shift depends on $P_m/P_S$ sign, hereinafter we chose $P_m/P_S > 0$ allowing for experimental result [13]. Notice, that the choice of $P_m/P_S$ sign leads to $P_V < 0$ in Figs.2, 3. The latter is followed from Eq. (18), because the equilibrium $P_V$ at $E_0 = 0$ has to be negative at $E_m \sim P_m > 0$ (see Eq.(22)).

The similarity of the calculated and observed hysteresis loops speaks in favour of the statement that the mechanism of self-polarization phenomenon in thin ferroelectric films could be mismatch effect. Dependence $\bar{P}(E_0)$ at $P_m = 0$ for the thinnest film ($h = 8$) has the form characteristic for paraelectric phase. This confirms the above-made statements, that critical thickness is close to $h = 8$.

One can see that for the loops depicted in Fig.4a ($0 < P_m/P_S < 1$) the coercive field increases with film thickness increase. Really this field can be found from expression (18) under the condition $d^2\Delta G/dP_V^2\big|_{P_V = P_{VC}} = 0$, that leads to the equation

$$A_m + 3B_m P_{VC}^2 + 2D_m P_{VC} = 0 .$$ (26)

The quadratic Eq.(26) can be easily solved, namely

$$P_{VC}(T,l) = -\frac{D_m(l)}{3B_m(l)} \pm \sqrt{\left(\frac{D_m(l)}{3B_m(l)}\right)^2 - \frac{A_m(T,l)}{3B_m(l)}} .$$ (27a)

Substitution $P_V = P_{VC}$ into Eq.(24) yields the exact expression for the coercive field $E_{0C}(T,l)$:

$$E_{0C}(T,l) = E_m(l) - P_{VC}^2 \left[D_m(l) + 2B_m(l)P_{VC}\right] .$$ (27b)

Keeping in mind the inequalities (15), we suppose that mismatch-induced polarization $P_m$ satisfies the conditions

$$\left(\frac{P_m}{P_S}\right)^2 < \frac{3(1+\Lambda)^2}{\theta}, \qquad \left(\frac{P_m}{P_S}\right)^2 << \frac{4h^2(1+\Lambda)^2}{\theta^2} .$$ (28)

Thus the Eq. (27a), when neglected the term $\left(D_m(l)/3B_m(l)\right)^2$, could be approximated as following:

$$P_{VC}(T,l) \approx -\frac{D_m(l)}{3B_m(l)} \pm \sqrt{-\frac{A_m(T,l)}{3B_m(l)}} .$$ (29)

When substituted (29) into (27b), with the accuracy of $O\left(D_m(l)/3B_m(l)\right)^2$ one obtains that in linear approximation on $P_m/P_S$

$$E_{0C}(T,l) \approx \pm\frac{2}{3}\sqrt{-\frac{A_m^3(T,l)}{3B_m(l)}} + E_m(l) - D_m(l)\frac{A_m(T,l)}{3B_m(l)} .$$ (30)

Without mismatch effect ($P_m = 0$) $D_m = 0$, $E_m = 0$ so that

$$P_{VC}(T,l) \approx \pm\sqrt{-\frac{A_m(T,l)}{3B_m(l)}} , \qquad E_{0C}(T,l) \approx \pm\frac{2}{3}\sqrt{-\frac{A_m^3(T,l)}{3B_m(l)}} .$$ (31)

Substitution of $A_m(T,l)$ from Eq.(19) into (31) yields the following expression for coercive field $E_{0C}$:

$$E_{0C}(P_m=0,T,h) = \pm \frac{2}{3\sqrt{3}} E_S \sqrt{\left(1 - \frac{T}{T_C^*} - \frac{1}{\theta\, h(1+\Lambda)}\right)^3}, \qquad \frac{T}{T_C^*} < 1. \tag{32}$$

One can see that the third term decreases with $h$ increase, so that $E_{0C}(P_m=0)$ increases with thickness increase. For the parameters used in Fig.4a ($\theta = 0.01$, $\Lambda = 10$, $T = 0$) analytical expression (32) gives $E_{0C}(P_m=0,h=40)/E_S = 0.25$ and $E_{0C}(P_m=0,h=20)/E_S = 0.13$, the ratio of these two coercive fields are in a good agreement with the results of numerical calculations depicted in Fig.4a (see curves 1). At $h \to \infty$ the Eq.(32) transforms into conventional expression for coercive field of bulk materials. The formulae (32) makes it possible to calculate $h_{cr}$ or $T_{cr}$ that corresponds to $E_{0C} = 0$, namely:

$$h_{cr}(P_m=0,T) = \frac{1}{\theta\left(1+\Lambda\right)\left(1-T/T_C^*\right)}, \quad T_{cr}(P_m=0,h) = T_C^*\left(1 - \frac{1}{\theta\left(1+\Lambda\right)h}\right). \tag{33}$$

Since $E_{0C} = 0$ at $P_m = 0$ corresponds to paraelectric phase, $h_{cr}$ or $T_{cr}$ has to be the critical thickness or temperature of ferroelectric-paraelectric phase transition and Eqs.(33) coincide completely with the corresponding expressions in [10], [11]. Note, that at $P_m \neq 0$ and $h \leq 8$ the hysteresis loop width $\Delta_H(T,l) \approx 4/3\sqrt{-A_m^3(T,l)/3B_m(l)}$ (see Eq.(30)) is close to zero. However, the absence of the width does not speak about paraelectric phase for all $P_m/P_S$ ratios. It may be related to impossibility of switching the large $E_m$ field in a thin film.

Let us now consider the thickness dependence of coercive field given by Eq.(30), that takes into account mismatch effect contribution in linear approximation on $P_m/P_S$. Allowing for $E_m \sim 1/h$, $D_m \sim 1/h$ (see Eqs.(21), (22)) the difference $\Delta E_{0C}$ between coercive fields at $P_m \neq 0$ and at $P_m = 0$ have to be inversely proportional to the film thickness $h$. One can see from the Fig.4a that the ratio $\Delta E_{0C}(h=40)/\Delta E_{0C}(h=20) \approx 0.5$, i.e. it is really close to inverse thickness ratio.

For large enough $P_m$ ($P_m/P_S > 1$) one has to take into account nonlinear on $P_m/P_S$ terms in $A_m(T,l)$ and $E_m(l)$ which do not change the dependence of these coefficients on film thickness $h$ (see Eqs.(19), (22)). Therefore the influence of mismatch effect decreases with the film thickness increase. Since the measurements of coercive field are widely used, the theoretical forecast about the dependence of $E_{0C}$ on thickness $l$ and mismatch parameter $P_m$ could be checked and used for our model justification.

It is evident that the field $E_m(l)$ plays the role of a bias field. It is known that biased ferroelectric hysteresis loops are often observed experimentally (see Fig.4b and [14] - [15]). Using the experimental value of bias field $E_0$ that makes the loop symmetrical at $E_0 = E_m$, one can obtain the quantity $P_m/P_S$ with the help of the cubic equation (22). Allowing for the condition $\theta << 1$, this equation has the solution $P_m/P_S \approx 2h\,\theta\left(1+\Lambda\right)E_m / E_S$. The extrapolation length $\Lambda$ should be obtained from the independent measurements, e.g. from the pyroelectric current spectrum [16]. The dependence of the ratio $E_m(h)/E_S$ over film thickness $h$ is represented in Fig.5. It is seen that the internal field increases with thickness decrease and $E_m(h)/E_S \sim 1/h$ (see inset to Fig.5). One can see from Eq.(24), that the polarization and other properties have to be dependent on difference $E_0 - E_m(l)$, so that screening of external field $E_0$ by internal field $E_m$ could be expected.

## 4. DIELECTRIC SUSCEPTIBILITY AND PHASE DIAGRAM

The linear dielectric susceptibility $\chi = dP_V/dE_0|_{E_0=0}$ could be obtained from (24), namely

$$\chi(T,l) = \frac{1}{A_m + 3B_m P_V^2 + 2D_m P_V}\bigg|_{E_0 \to 0}. \tag{34}$$

Let us find the temperature of susceptibility maximum or divergence - $T_m(l)$ - from the condition $d\chi(T,l)/dT = 0$. The derivative $dP_V(T,l)/dT$ obtained from the Eq.(24) corresponds to the equilibrium condition $P_V(T_m,l)(E_0 - E_m(l)) \geq 0$. After some elementary transformations (see Appendix A) one obtains that the maximum of static susceptibility must satisfy the conditions:

$$P_V^2(T_m,l) = \frac{A_m(T_m,l)}{3 B_m(l)}, \tag{35a}$$

$$\chi(T_m,l) = \frac{1}{2(3 B_m P_V^2(T_m,l) + D_m(l) P_V(T_m,l))}, \tag{35b}$$

$$4 P_V^3(T_m,l) + \frac{D_m(l)}{B_m(l)} P_V^2(T_m,l) = \frac{E_0 - E_m(l)}{B_m(l)}. \tag{35c}$$

It should be underlined, that Eq.(35c) reflects the influence of external electric field $E_0$ on the value of $T_m$. To obtain the expression for $T_m(l,E_0)$ let us make some simplifications in Eq.(35c). Namely, having used (15), (28), one could neglect the term $D_m(l)P_V^2(T_m,l)/B_m(l)$ in (35c) and obtain that

$$T_m(l,E_0) = T_C^* \left[ 1 - \frac{1}{\theta\, h(l)(1+\Lambda)} + 3\left(\frac{E_0 - E_m(l)}{4 E_S}\right)^{2/3} \right]. \tag{36}$$

Let us put $E_0 = 0$ in Eqs.(22), (36) and obtain that under zero-field conditions

$$T_m(h) = T_C^* \left[ 1 - \frac{1}{\theta\, h(1+\Lambda)} + \frac{3}{4}\left(\frac{P_m/P_S}{\theta\, h(1+\Lambda)}\right)^{2/3} \right]. \tag{37}$$

The thickness dependence of the maximum susceptibility temperature $T_m(l)$ is depicted in Fig.6, which could be regarded as a phase diagram, allowing for $T_m(l)$ used to be considered as the temperature of ferroelectric-paraelectric phase transition. One can see the essential difference between the phase diagram for $0 < P_m/P_S < 1$ and $P_m/P_S \geq 1$ (compare the Figs.6a and 6b). It follows from Fig.6a, that the critical thickness that corresponds to $T_m = 0$ decreases with mismatch effect increase, the dependence $(T_m - T_C^*) \sim 1/h$ is in agreement with previous results at $P_m = 0$ [10]. Contrary to this at $P_m/P_S = 2$ the temperature $T_m \neq 0$ even at $l = 4\, l_d$, i.e. for ultrathin (close to monolayer) film (see curve 3 in the inset to Fig.6b). The further increase of $P_m/P_S$ ratio made $T_m/T_C^* > 1$ for wide region of the film thickness (see curves 4, 5 in Fig.6b). Therefore there is no possibility to obtain $T_m = 0$ for $P_m/P_S > 1$ at any small thickness of a film and so there is no thickness induced ferroelectric-paraelectric phase transition. This could explain the conservation of ferroelectric phase in practically monolayer films, observed earlier (see e.g. [17]). The value $P_m/P_S = 1$ may be considered as a boundary between these two types of behaviour - with and without thickness induced transition. However as it follows from the curve 2 in Fig.6b, that for the monolayer film $T_m \approx 0$, so for $P_m/P_S = 1$ the phase transition could exist. For the case $0 \leq P_m/P_S \leq 1$ one can expect the anomalies in dielectric permittivity at critical temperature or critical thickness.

The maximum linear static susceptibility could be obtained from (35b) and (35c) under the conditions of (28) validity as follows:

$$\chi(T_m,l) \approx \frac{1}{3 \sqrt[3]{B_m E_m(l)^2/2}}. \tag{38}$$

One can see from (38) that susceptibility (34) diverges at $T = T_m(l)$ only if $E_m = 0$, i.e. $P_m = 0$ (see Eq.(22)). In such case $A_m(T_m,l) = 0$ because $A_m(T_m,l) \approx 3/2 \sqrt[3]{B_m E_m(l)^2/2}$ at $E_0 = 0$. Moreover at $P_m \neq 0$ susceptibility $\chi(T,l)$ has only maximum at $T = T_m(l)$, which diffuses under $P_m$ absolute value increasing (see Fig.7). It follows from Eqs.(38), (22) that the height of the maximum has the view $\chi(T_m, h) \sim h^{2/3} (P_m/P_S)^{-2/3}$, so that the maximum becomes sharper and higher with the film thickness increase and mismatch polarization decrease. The obtained power law with exponent 2/3 for this behaviour can be checked with the help of experimental data for the film susceptibility dependence on the temperature, film thickness and the type of the substrate. The analytical expression of the depicted in Fig.7 shift of $T_m$ in the dependence on the film thickness and mismatch effect value is given by Eq.(37), namely the shift increases as $C_1 + C_2 (P_m/P_S)^{2/3}$ at fixed thickness and decreases as $C_3 - C_4/h + C_5/h^{2/3}$ at fixed $P_m$ value. Here $C_{1,2}$ and $C_{3,4,5}$ are constants independent on $P_m$ and $h$ values respectively. The observation of the $T_m$ shift can be also useful to check the validity of proposed model.

It should be noted that we calculated all the Figs.2-7 on the basis of exact formulas, while approximate expressions, e.g. (36)-(38), were derived for the sake of illustrations.

## DISCUSSION

The mismatch effect considered in this paper is related to mechanical strain tensor $u_{xx}$ originated from the difference in a substrate and a film lattice constants, thermal coefficients and growth imperfections. The latter depends on technological process of a film fabrication while the others are defined by the pair substrate-film. The mechanical strain influences the electric polarization dependent bulk part of free energy via electrostriction (see Eqs.(6), (8)) and via piezoelectric effect on the surface (see Eqs.(4), (5)). The latter effect appears even in cubic lattices because of the absence of inversion centre in the vicinity of the surface. Because the inversion disappears in $z$ direction only, while it exists in $x$ and $y$ ones, the only non-zero piezoelectric coefficient related to $u_{xx}$ can be $d_{xxz} \equiv d$. As a result permanent electric polarization on the surface $P_z \equiv P_m$ arises (see Eq.(7)), while nothing of this kind exists in $x$ or $y$ directions. Because of the fact we supposed that mismatch effect caused the film self-polarization phenomenon. The shift of hysteresis loops and coercive field asymmetry $\Delta E_{0C}$ obtained in this paper is known to be the characteristic feature observed experimentally in self-polarized film (see Fig.4b).

Since $\Delta E_{0C}$ and coercive field itself are defined essentially by $E_m$ field, let us estimate its value. To do this we took the following reasonable values of the parameters: $d \sim (10^{-5} \div 10^{-6}) CGSE$, $\delta \sim (10^{-14} \div 10^{-16}) cm^2$, $U_m \sim 10^{-2}$, $\theta \sim 5 \cdot 10^{-2}$, $l_d \sim 10^{-8} cm$, $\mu \sim (5 \div 0.5) \, 10^4 \, din/cm^2$, so $P_m = U_m/d \sim (10^3 \div 10^4) CGSE$ and

$E_m \approx \dfrac{4\pi P_m (l_d/l)}{(1+\lambda/l_d)} \sim \dfrac{l_d/l}{(1+\lambda/l_d)} (10^4 \div 10^5) CGSE \sim \dfrac{l_d/l}{(1+\lambda/l_d)} 3 \cdot (10^3 \div 10^4) kV/cm$. For $l/2l_d \sim 10$

and $\lambda/l_d \sim 10$ one obtains that $E_m \sim (30 \div 300) \, kV/cm$. It is which is in reasonable agreement with the experimental value $E_{0C} \sim 200 \, kV/cm$ obtained for thick enough film (see Fig.4b). Note, that obtained $E_m$ value shows, that this mismatch-induced internal field is really able to induce self-polarization and to screen (at least partially) external field $E_0$ in thin ferroelectric films.

On the other hand the obtained here theoretical forecast of $\Delta E_{0C} \sim P_m/(hP_S)$ for $|P_m|/P_S < 1$ should be compared with experimentally observed size effect of coercive field asymmetry (if any) and its dependence on the type of substrate and technological process for self-polarized films. However, it should be noted, that the proposed explanation of self-polarization phenomenon can be considered as preliminary one, because more detailed calculations of this phenomenon have to include the calculations on the basis of free energy with three components of polarization $P_x$, $P_y$ and $P_z$. These calculations are in progress now.

The observation of the other theoretical forecasts, e.g. $\chi(T_m,h) \sim h^{2/3}\left(P_S/P_m\right)^{2/3}$, as well as the peculiar dependence of $\chi(T_m,h,P_m)$ (see Fig.7) could be useful to check the validity of the proposed model, when size effects are related to thickness induced phase transitions, namely, for the cases $\left|P_m\right|/P_S < 1$ (i.e. when the polarization induced by mismatch effect on the surface is smaller than the bulk one). The latter is also valid at $P_m = 0$, because extrapolation length $\lambda > 0$ as follows from Eq.(7) because $\delta > 0$, $\mu > 0$. Positive sign of extrapolation length makes it impossible to explain the reasons of ferroelectric phase stability in some monolayer films (see e.g. [17]). Our consideration of surface polarization induced by the large enough mismatch effect (the cases $P_m/P_S > 1$) had shown that the physical mechanism of ferroelectric phase conservation up to one layer film is the mismatch effect. The experimental confirmation of this statement by different choice of the pair film-substrate is extremely desirable.

## APPENDIX A

Let us find the temperature $T_m(l)$ of susceptibility maximum or divergence from Eq.(24) and the condition $d\chi(T,l)/dT = 0$. Taking into account that only $A_m$ and $P_V$ depend on temperature $T$ we derive that:

$$\frac{d\chi(T,l)}{dT} = -\frac{dA_m/dT + \left(6B_m P_V + 2D_m\right)dP_V/dT}{\left(A_m + 3B_m P_V^2 + 2D_m P_V\right)^2} = 0, \qquad (A.1)$$

$$\frac{d\left(A_m P_V + B_m P_V^3 + D_m P_V^2\right)}{dT} \equiv P_V\frac{dA_m}{dT} + \left(A_m + 3B_m P_V^2 + 2D_m P_V\right)\frac{dP_V}{dT} = \frac{d(E_0 - E_m)}{dT} \equiv 0.$$

Directly from the system (A.1) one obtains that

$$\frac{dP_V}{dT} = -\frac{dA_m}{dT}\frac{1}{6B_m P_V + 2D_m},$$
$$\frac{dP_V}{dT} = -P_V\frac{dA_m}{dT}\frac{1}{A_m + 3B_m P_V^2 + 2D_m P_V}. \qquad (A.2)$$

The left hand-sides of Eqs.(A.2) are identical, thus from the identity of the right hand-sides we obtain the following equation for $P_V(T_m,l)$:

$$\frac{1}{6B_m P_V + 2D_m} = \frac{P_V}{A_m + 3B_m P_V^2 + 2D_m P_V}. \qquad (A.3)$$

From the Eq.(A.3) we obtain $A_m + 3B_m P_V^2 + 2D_m P_V = P_V\left(6B_m P_V + 2D_m\right)$ and so:

$$P_V^2(T_m,l) = \frac{A_m(T_m,l)}{3B_m(l)}. \qquad (A.4)$$

Having substituted (A.4) in the form $A_m(T_m,l) = 3B_m(l)P_V^2(T_m,l)$ into (24) and (34), we immediately obtain Eqs.(35c) and (35b) correspondingly. In contrast, having substituted (A.4) directly into (24), one obtains the equation for $A_m(T_m,l)$, namely:

$$\pm 4B_m(l)\left(\frac{A_m(T_m,l)}{3B_m(l)}\right)^{3/2} + \frac{A_m(T_m,l)D_m(l)}{3B_m(l)} = E_0 - E_m(l). \qquad (A.5)$$

In accordance with the inequalities (28) the second term in the left hand-side of Eq.(A.5) could be neglected and then $A_m(T_m,l) \approx 3/2\sqrt[3]{B_m\left(E_0 - E_m(l)\right)^2/2}$. The formulas (37), (38) were derived from the latter expression and Eq.(19).

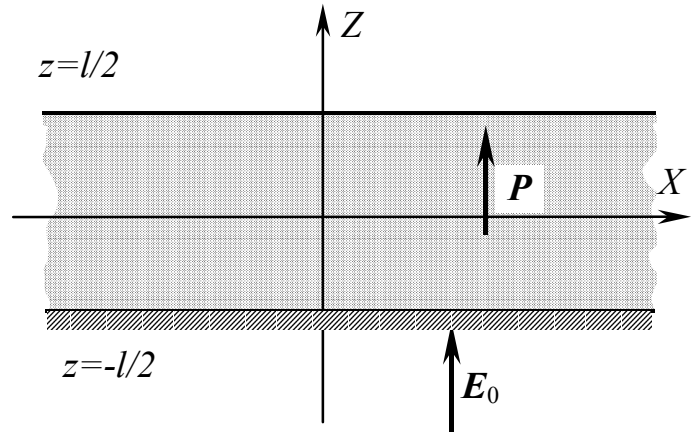

Figure 1. The scheme of the film on the substrate (*////////*).

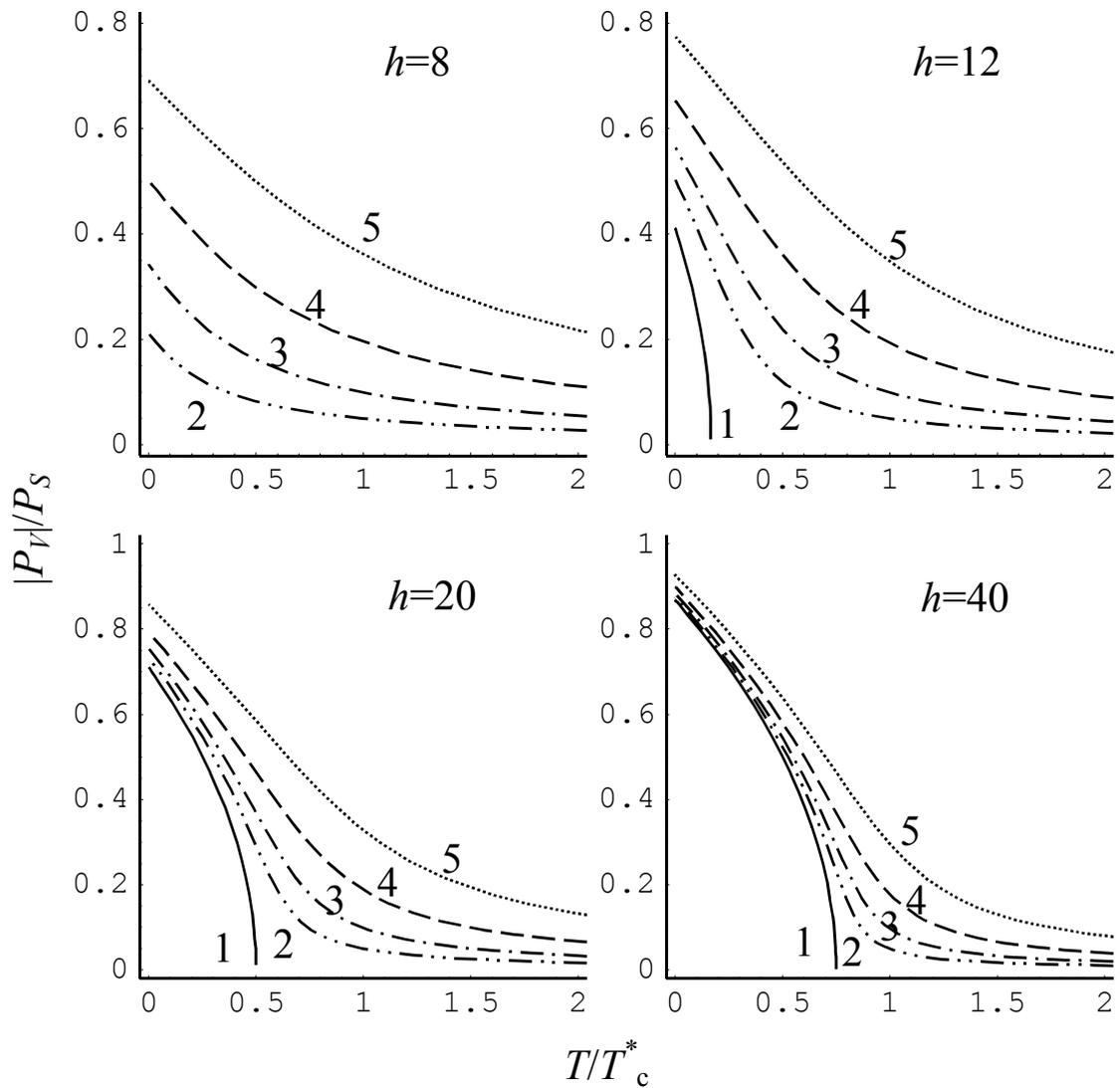

Figure 2. The temperature dependence of the polarization for $\theta = 0.01$, $\Lambda = 10$, $E_0 = 0$ and different $P_m/P_S$ values: 0 (curves 1), 0.1 (curves 2), 0.2 (curves 3), 0.4 (curves 4), 0.8 (curves 5).

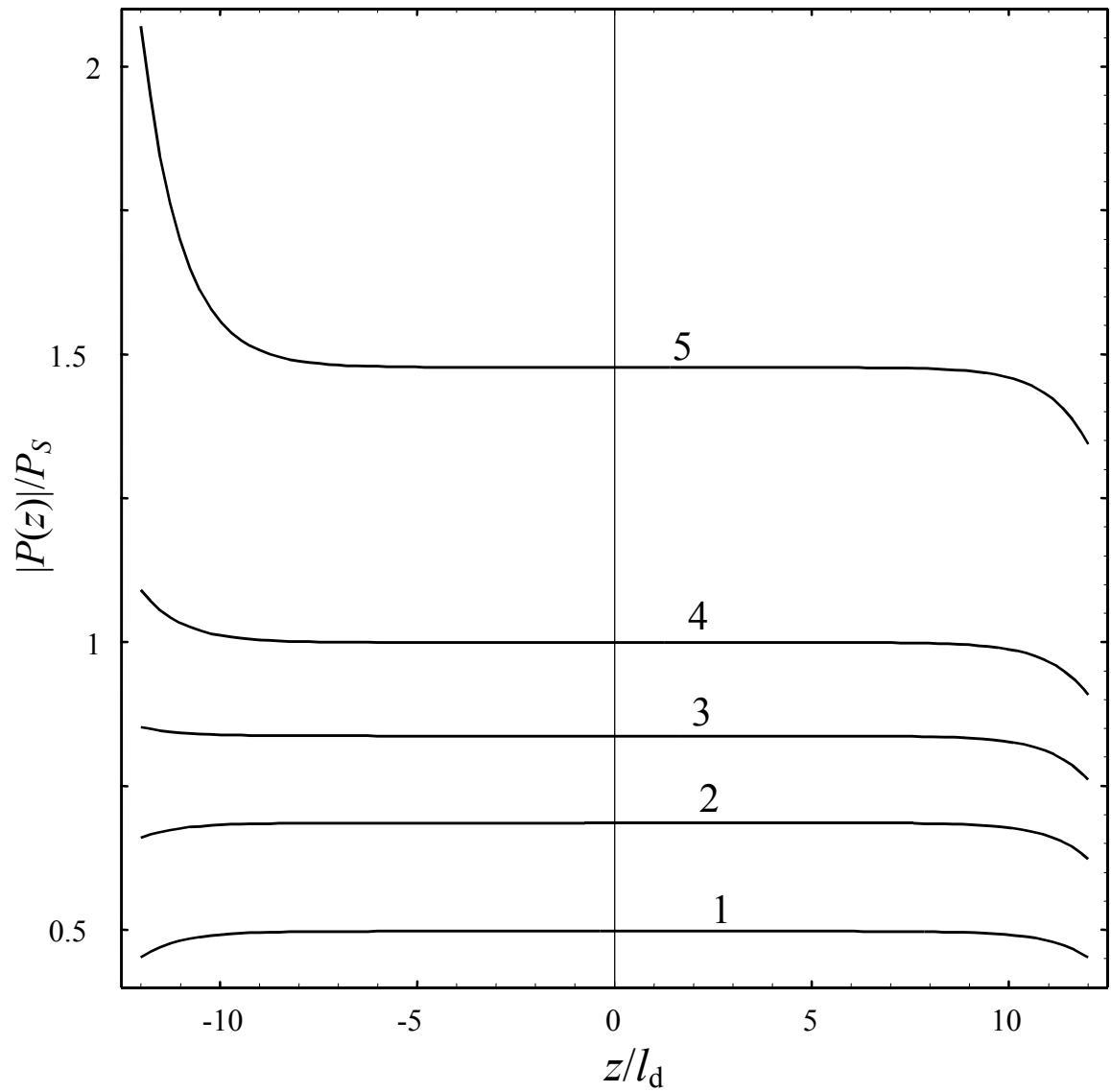

Figure 3. The distribution of the polarization over coordinate $z$ inside the film for $\Lambda = 10$ and different $P_m/P_S$ values: $P_m/P_S = 0$, 0.4, 1, 2, 8 (curves 1, 2, 3, 4, 5 respectively).

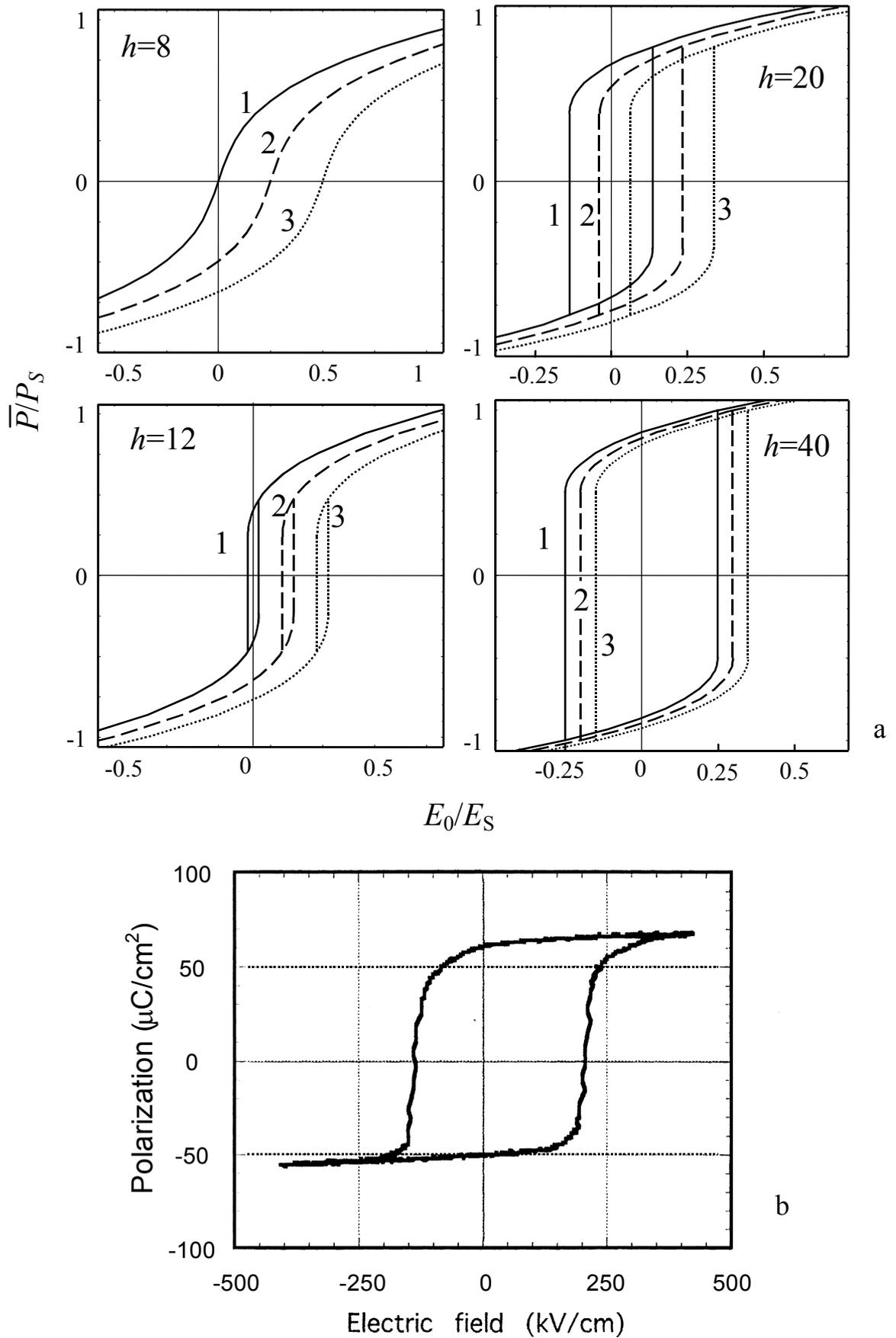

Figure 4. The typical forms of hysteresis loops $P(E_0)$:
a – theoretical calculation for $\theta = 0.01$, $\Lambda = 10$, $T = 0$ and different $P_m/P_S$ values: 0 (curves 1),
0.4 (curves 2), 0.8 (curves 3); b – loop of the PZT film on MgO substrate [13].

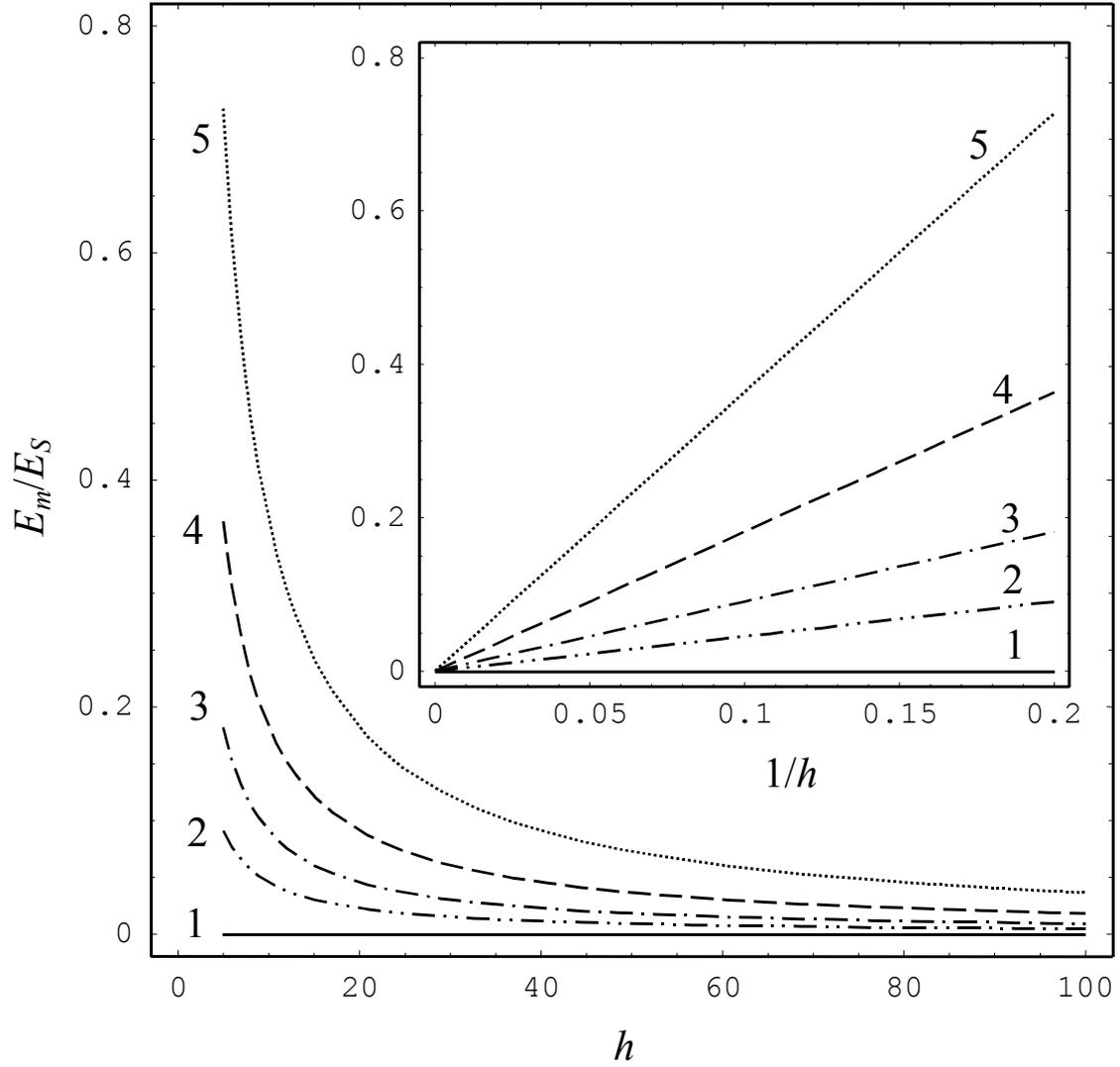

Figure 5. The dependence of the dimensionless internal electric field $E_m$ over film thickness $h$ for $\theta = 0.01$, $\Lambda = 10$ and different $P_m/P_S$ values: 0 (curve 1), 0.1 (curve 2), 0.2 (curve 3), 0.4 (curve 4), 0.8 (curve 5).

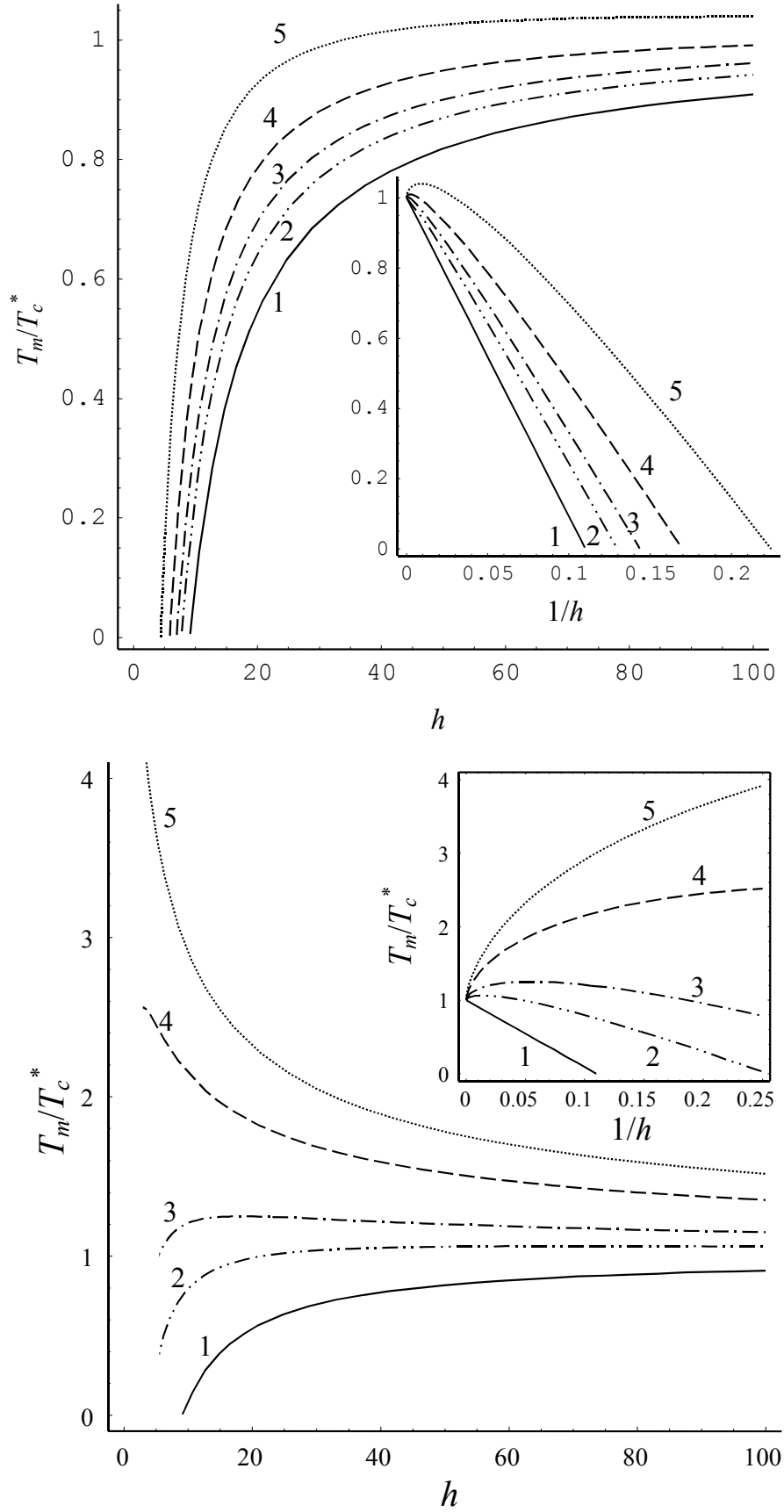

Figure 6. The dependence of the susceptibility maximum temperature $T_m$ over film thickness $h$ for $\theta = 0.01$, $\Lambda = 10$, $E_0 = 0$ and different $P_m/P_S$ values:
a − 0 (curve 1), 0.1 (curve 2), 0.2 (curve 3), 0.4 (curve 4), 0.8 (curve 5).
b − 0 (curve 1), 1 (curve 2), 2 (curve 3), 5 (curve 4), 8 (curve 5).

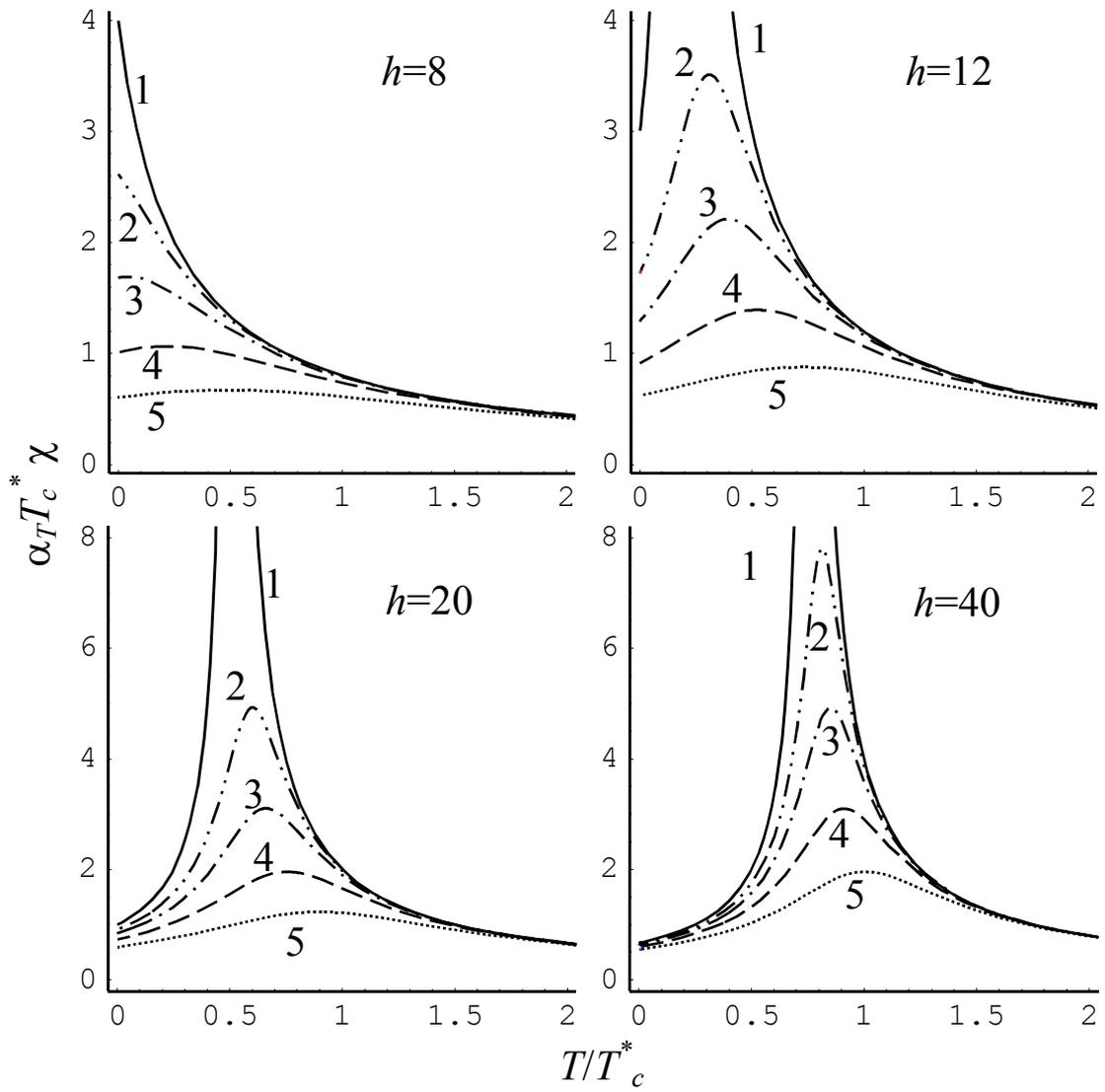

Figure 7. The temperature dependence of the averaged susceptibility for θ = 0.01, Λ = 10, $E_0 = 0$ and different $P_m/P_S$ values: 0 (curves 1), 0.1 (curves 2), 0.2 (curves 3), 0.4 (curves 4), 0.8 (curves 5).